\documentclass{article}
\usepackage{amsfonts, amsmath, amsthm, amssymb, graphicx, latexsym, slashbox, natbib, bm, color,enumerate}

\usepackage[left=1in,top=1in,right=1in,bottom=1in,nohead,paperwidth=8.5in, paperheight=11in]{geometry} 
\title{\textbf{Robust Sparse Canonical Correlation Analysis}}
\author{Ines Wilms and Christophe Croux}

\date{}

\begin{document}
\maketitle

\textbf{Abstract}:
Canonical correlation analysis (CCA) is a multivariate statistical method which describes the associations between two sets of variables. The objective is to find linear combinations of the variables in each data set having maximal correlation. This paper discusses a method for Robust Sparse CCA. Sparse estimation produces canonical vectors with some of their elements estimated as exactly zero. As such, their interpretability is improved. We also robustify the method such that it can cope with outliers in the data. To estimate the canonical vectors, we convert the CCA problem into an alternating regression framework, and use the sparse Least Trimmed Squares estimator. We illustrate the good performance of the Robust Sparse CCA method in several simulation studies and two real data examples.

\bigskip



\section{Introduction}

Canonical correlation analysis (CCA), introduced by \cite{Hotelling36}, identifies and quantifies the associations between two sets of variables. CCA searches for linear combinations, called \textit{canonical variates}, of each of the two sets of variables having maximal correlation.  The coefficients of these linear combinations are called the \textit{canonical vectors}. The correlations between the canonical variates are called the \textit{canonical correlations}. For more information on canonical correlations analysis, see e.g. \citeauthor{Johnson98} (\citeyear{Johnson98}, Chapter 10).

\textit{Sparse} canonical vectors are canonical vectors with some of their elements estimated as exactly zero. The canonical variates then only depend on a subset of the variables, those corresponding to the non-zero elements of the estimated canonical vectors. Hence, the canonical variates are easier to interpret, in particular for high-dimensional data sets. Examples of CCA for high-dimensional data sets can be found in, for example, genetics \citep{Gonzalez08, Prabhakar2012, CruzCano} and machine learning \citep{Sun2011}.

Different approaches for sparse CCA have been proposed in the literature. \cite{Parkhomenko09} use a sparse singular value decomposition to derive sparse singular vectors. \cite{Witten09} develop a penalized matrix decomposition, and show how to apply it for sparse CCA. \cite{Waaijenborg08}, \cite{Lykou10}, and \cite{Wilms13} convert the CCA problem into a penalized regression framework to produce sparse canonical vectors. All these methods are not robust to outliers. A common problem in multivariate data sets, however, is the frequent occurrence of outliers. Therefore, the possible presence of outliers should be taken into account.

Several \textit{robust CCA} methods have been introduced in the literature. \cite{Karnel91} considers robust CCA using an M-estimator of multivariate location and scatter; \cite{Dehon02} use the minimum covariance determinant (MCD, \citealp{Rousseeuw85}) estimator. Asymptotic properties for CCA based on robust estimators of the covariance matrix are discussed in \cite{Taskinen06}.  \cite{Filzmoser00} use a robust alternating regression approach, following the approach of \cite{wold1968}, to obtain the first canonical variates. \cite{Branco05} extend the method of \cite{Filzmoser00} to obtain all canonical variates. 
CCA can also be considered as a prediction problem, where the canonical variates obtained from the first data set serve as optimal predictors for the canonical variates of the second data set, and vice versa.
As such, \cite{Adrover15} use a robust M-scale to evaluate the prediction quality, whereas the approach of  \cite{Kudraszow11} is based on a robust estimator for the multivariate linear model. None of these methods, however, are sparse. 

This paper proposes a CCA method that is sparse and robust at the same time. As such, we deal with two important topics in applied statistics: sparse model estimation and the presence of outliers in the data. We consider CCA from a predictive point of view. We use an alternating robust, sparse regression framework to sequentially obtain the canonical variates. We obtain sparse canonical vectors that are resistant to outlying observations by using the sparse Least Trimmed Squares (sparse LTS) estimator of \cite{Alfons13}. We show that the Robust Sparse CCA method has a clear advantage over standard CCA and Sparse CCA by means of a simulation study. The advantage of Robust Sparse CCA over  Robust CCA  is twofold. First, Robust Sparse CCA provides well interpretable canonical vectors since some of the elements of the canonical vectors are estimated as exactly zero. Secondly, when the number of variables becomes large compared to the sample size, the estimation accuracy of Robust Sparse CCA will be better. Moreover, when the largest number of variables of both data sets exceeds the sample size, Robust Sparse CCA can still be performed.

The remainder of this article is organized as follows. Section 2 considers the CCA problem from a predictive point of view. Section 3 discusses the Robust Sparse CCA algorithm. Section 4 presents simulation results where we compare Robust Sparse CCA  to standard CCA, Robust CCA and Sparse CCA. Two real data examples are discussed in Section 4. Section 5 concludes.

\section{CCA based on alternating regressions}
We consider the CCA problem from a predictive point of view, as proposed by \cite{Brillinger75} and \cite{Izenman75}. Given a sample of $n$ observations ${\bf x_i} \in \mathbb{R}^{p}$ and  ${\bf y_i} \in \mathbb{R}^{q}$ ($i=1,\ldots,n$). The two data matrices are denoted as ${\bf X}=[{\bf x_1},\ldots,{\bf x_n}]^{T}$ and ${\bf Y}=[{\bf y_1},\ldots,{\bf y_n}]^{T}$.
The estimated canonical vectors are collected in the columns of the matrices $\widehat{{\bf A}} \in \mathbb{R}^{p \times r}$ and $ \widehat{{\bf B}} \in \mathbb{R}^{q \times r}$. Here $r$ is the number of canonical vectors. 
The  columns of the matrices $\bf{X}\widehat{{\bf A}}$ and $\bf{Y}\widehat{{\bf B}}$ contain the estimates of the realizations of the  canonical variates, and we denote their $j^{th}$ column by $ \bf{\hat u}_j$ and $ \bf{\hat v}_j$, for $1 \leq j  \leq r$. 
The objective function defining the canonical vector estimates is
\begin{equation}
(\widehat{{\bf A}}, \widehat{{\bf B}}) = \underset{({\bf A}, {\bf B}) }{\operatorname{argmin}} \ \sum_{i=1}^{n} || {\bf A}^T{\bf x_i} - {\bf B}^T {\bf y_i} ||^2. \label{objfunction}
\end{equation}
The objective function in \eqref{objfunction} is minimized under the restriction that each canonical variate $\bf{\hat u}_j$ is uncorrelated with the lower order canonical variates $\bf{\hat u}_k$, with $1 \leq k < j \leq r$. Similarly for the canonical vectors within the second set of variables. For identification purpose, a normalization condition requiring the canonical vectors to have unit norm is added.  
Typically, the canonical vectors are obtained by an eigenvalue analysis of a certain matrix involving the inverses of sample covariance matrices. But if $n<\text{max}(q,p)$, these inverses do not exist.

We estimate the canonical vectors with an alternating regression procedure using the Least Squares estimator \citep{wold1968}. 
Indeed, if the matrix $\bf{A}$ in \eqref{objfunction} is kept fixed, the matrix $\bf{B}$ can be obtained from a Least Squares regression of the canonical variates on $\bf{y}$ (and vice versa for estimating $\bf{A}$ keeping $\bf{B}$ fixed). The Least Squares estimator, however, is not sparse, nor robust to outliers. Therefore, we replace it by the sparse Least Trimmed Squares (sparse LTS) estimator \citep{Alfons13}. The sparse LTS estimator is a sparse version of the Least Trimmed Squares estimator. Sparse model estimates are obtained by adding an $L_1$ penalty to the LTS objective function, similar as for the lasso regression estimator \citep{Tibshirani96}. The sparse LTS estimator can be applied to high-dimensional data, where the sample size exceeds the number of variables and is therefore as well a regularized version of LTS.
 
\section{Robust Sparse alternating regression algorithm}
We use a sequential algorithm to derive the canonical vectors. 
\bigskip

{\it First canonical vector pair.} Denote the first canonical vector pair by $({ \bf A_1 }, { \bf B_1 } )$. Assume that the value of ${\bf A_1}$ is known. Denote the vector of squared residuals by ${\bf r}^2({{ \bf B}_1})= (r_1^2,\ldots,r_n^2)^T$, with $r_i^2 = ({\bf A_1}^T{\bf x_i} - {\bf B_1}^T {\bf y_i})^2,  i=1,\ldots,n$. The estimate of $\bf B_1$ is obtained as
\begin{equation}
\widehat{ \bf B}_1|{ \bf A_1}  = \underset{{\bf B_1}}{\operatorname{argmin}} \sum_{i=1}^{h} \left( {\bf r}^2({{ \bf B}_1}) \right )_{i:n} + \lambda_{B_1} \sum_{j=1}^{p} |{{ \bf B}_1}_j|  , \label{objfunction_B1}
\end{equation}
where $\lambda_{B_1}>0$ is a sparsity parameter, ${{ \bf B}_1}_j$ is the $j^{th}$ element, $j=1,\ldots,p$, of the first canonical vector  ${\bf B}_1$,  and  $\left( {\bf r}^2({{ \bf B}_1}) \right )_{1:n} \leq \ldots \leq \left( {\bf r}^2({{ \bf B}_1}) \right )_{n:n}$ are the order statistics of the squared residuals. The canonical vector $\widehat{ \bf B}_1$ is normed to length 1. The solution to \eqref{objfunction_B1} equals the sparse LTS estimator with ${\bf X}{\bf A_1}$ as response and ${\bf Y}$ as predictor. 
Regularization by adding a penalty term to the objective function is necessary since the design matrix ${\bf Y}$ can be high-dimensional. 
As such, we get a robust sparse estimate $\widehat{ \bf B}_1$. 

Analogously, for a fixed value $\bf B_1$, denote the vector of squared residuals by ${\bf r}^2({{ \bf A}_1})= (r_1^2,\ldots,r_n^2)^T$, with $r_i^2 = ({\bf B_1}^T{\bf y_i} - {\bf A_1}^T {\bf x_i})^2, i=1,\ldots,n$. The sparse LTS regression estimate of $\bf A_1$ with ${\bf Y}{\bf B_1}$ as response and ${\bf X}$ as predictor, is
\begin{equation}
\widehat{ \bf A}_1|{ \bf B_1}  = \underset{{\bf A_1}}{\operatorname{argmin}} \sum_{i=1}^{h} \left( {\bf r}^2({{ \bf A}_1}) \right )_{i:n} + \lambda_{A_1} \sum_{j=1}^{q} |{{ \bf A}_1}_j|, \label{objfunction_A1}
\end{equation}
where $\lambda_{A_1}>0$ is a sparsity parameter, ${{ \bf A}_1}_j$ is the $j^{th}$ element, $j=1,\ldots,q$ of the first canonical vector  ${\bf A}_1$, and $\left( {\bf r}^2({{ \bf A}_1}) \right )_{1:n} \leq \ldots \leq \left( {\bf r}^2({{ \bf A}_1}) \right )_{n:n}$ are the order statistics of the squared residuals. The canonical vector $\widehat{ \bf A}_1$ is normed to length 1.  

This leads to an alternating regression scheme, updating in each step the estimates of the canonical vectors until convergence. We iterate until the angle between the estimated canonical vectors in two successive iterations is smaller than some tolerance value $\epsilon$ (e.g. $\epsilon=10^{-3}$), and this for ${ \widehat{\bf A}}_1$ and  $\widehat{{ \bf B}}_1$. In the simulations we conducted, convergence was almost always reached.

\bigskip

{\it Higher order canonical vector pairs.} We use deflated data matrices to estimate the higher order canonical vector pairs (see e.g. \citealp{Branco05}).
For the second canonical vector pair, the deflated matrices are ${\bf X}^{*}$, the residuals of a column-by-column LTS regression of ${\bf X}$ on all lower order canonical variates,  ${\bf \hat{u}}_{1}$ in this case;  and ${\bf Y}^{*}$,  the residuals of a column-by-column LTS regression of ${\bf Y}$ on ${\bf \hat{v}}_{1}$. Since these regressions only involve a small number of regressors, they should not be regularized.

The second canonical variate pair is then obtained by alternating between the following regressions until convergence:
\begin{equation}
\widehat{ \bf B}_2^*|{ \bf A^*_2}  = \underset{{\bf B_2^*}}{\operatorname{argmin}} \sum_{i=1}^{h} \left( {\bf r}^2({{ \bf B}_2^*}) \right )_{i:n} + \lambda_{B_2^*} \sum_{j=1}^{p} |{{ \bf B}_2^*}_j|, \label{objfunction_B2}
\end{equation}
where ${\bf r}^2({{ \bf B}_2^{\star}})= (r_1^2,\ldots,r_n^2)^T$, with $r_i^2 = ({\bf A^*_2}^T{\bf x_i^{\star}} - {\bf B_2}^{\star T} {\bf y_i^{\star}})^2, i=1,\ldots,n$.
\begin{equation}
\widehat{ \bf A}_2^*|{ \bf B^*_2}  = \underset{{\bf A_2^*}}{\operatorname{argmin}} \sum_{i=1}^{h} \left( {\bf r}^2({{ \bf A}_2^*}) \right )_{i:n} + \lambda_{A_2^*} \sum_{j=1}^{q} |{{ \bf A}_2^*}_j|, \label{objfunction_A2}
\end{equation}
where ${\bf r}^2({{ \bf A}_2^{\star}})= (r_1^2,\ldots,r_n^2)^T$, with $r_i^2 = ({\bf B^*_2}^T{\bf y_i^{\star}} - {\bf A_2^{\star}}^T {\bf x_i^{\star}})^2, i=1,\ldots,n$.
The canonical vectors $\widehat{ \bf B}_2^*$ and $\widehat{ \bf A}_2^*$ are both normed to length 1.

Finally, the second canonical vector needs to be expressed as linear combinations of the columns of the original data matrices, and not the deflated ones. Since we want to allow for zero  coefficients in these linear combinations, a sparse approach is needed. To obtain a sparse $\widehat{ \bf A}_2$, we regress ${\bf \hat{u}}^*_{2}$ on ${\bf X}$ using the sparse LTS estimator, yielding the fitted values ${\bf \hat{u}}_{2}={\bf X}\widehat{ \bf A}_2$. To obtain a sparse $\widehat{ \bf B}_2$, we regress ${\bf \hat{v}}^*_{2}$ on ${\bf Y}$ using the sparse LTS estimator, yielding the fitted values ${\bf \hat{v}}_{2}={\bf Y}\widehat{ \bf B}_2$. 

The higher order canonical variate pairs are obtained in a similar way. We perform alternating sparse LTS regressions as in \eqref{objfunction_B2} and \eqref{objfunction_A2}, followed by a final sparse LTS step to retrieve the estimated canonical vectors $(\bf{\widehat{{}A}_k},\bf{\widehat{{}B}_k})$. Note that the regressions in \eqref{objfunction_B2} and \eqref{objfunction_A2} should be regularized, since the number of regressors equals $p$ or $q$, which could be larger than $n$, but sparsity is not really necessary.  

\bigskip
\textit{Initial value.} A starting value for ${\bf A_1}$ is required to start up the algorithm. Following \cite{Branco05}, we first obtain the first principal component of ${\bf Y}$, denoted ${\bf z_1}$. For this aim, we use the algorithm of \cite{RpcaPP} (available in the R package \verb+pcaPP+) which performs robust principal component analysis. We regress ${\bf z_1}$ on ${\bf X}$ using the LTS estimator, or the sparse LTS if the number of variables is larger than the sample size.  The estimated regression coefficient matrix of this regression is used as initial value for ${\bf A_1}$. 

\bigskip
\textit{Number of canonical variates to extract.} To decide on the number of canonical variates $r$ to extract, we use the maximum eigenvalue ratio criterion of \cite{An13}. We apply the Robust Sparse CCA algorithm and calculate the robust correlations $\hat{\rho}_1, \ldots,\hat{\rho}_{\text{rmax}} $, with $\text{rmax}=\text{min}(p,q)$. 
Each $\hat\rho_j$ is obtained by computing the correlation between ${\bf \hat{v}}_{j}$ and ${\bf \hat{u}}_{j}$ from the bivariate Minimum Covariance Determinant estimator with 25\% trimming. Let $\hat{k}_j = \hat{\rho}_j/\hat{\rho}_{j+1}$ for $j=1,\ldots,\text{rmax}-1$. We extract $r$ pairs of canonical variates, where $r=\text{argmax}_j \hat{k}_j$.

\bigskip
\textit{Reweighted sparse LTS.}  The sparse LTS estimator is computed with trimming proportion 25\%, so size of the subsample $h=\lfloor 0.75 n\rfloor$. To increase efficiency, we use a reweighting step afterwards. The reweighted sparse LTS is the lasso estimator computed from the observations not detected as outliers by the sparse LTS, i.e. having an absolute value of the standardized residuals smaller than or equal to the $98.75^{\text{th}}$ quantile of the standard normal distribution (see \citealp{Alfons13} for more detail).

\bigskip
\textit{Choice of the sparsity parameter.} The sparsity parameter controlling the penalization on the regression coefficient matrices is selected with the Bayesian Information Criterion (e.g. \citealp{Yin11}). We use a range of values for the sparsity parameter $\lambda$ and select the one with the lowest value of  
\begin{equation}\label{eq: BICbetaB}
BIC_{\lambda} = -2 \log L_{\lambda} + k_{\lambda} \log(n), \nonumber
\end{equation}
where $L_{\lambda}$ is the estimated likelihood, using sparsity parameter $\lambda$ and $k_{\lambda}$ is the number of non-zero estimated regression coefficients.

\section{Simulation Study}
We compare the performance of the Robust Sparse CCA method with (i) standard CCA, (ii) Robust CCA, and (iii) Sparse CCA.  The same algorithm is used for all 4 estimators, for ease of comparability. Robust CCA uses LTS instead of sparse LTS, and corresponds to the alternating regression approach of \cite{Branco05}. Standard CCA uses LS instead of LTS, Pearson correlations for computing the canonical correlations, and ordinary PCA for getting the initial values. Sparse CCA is like standard CCA, but used the lasso instead of LS for the multiple regressions.
\bigskip

Several simulation settings are considered. In the first simulation design (revised from \citealp{Branco05}), there is one canonical variate pair and the canonical vectors have a sparse structure. The canonical vectors are very sparse; each containing only one non-zero element. 
In the second design, there are two canonical variate pairs and the canonical vectors are non-sparse. 
In the third design, there are a lot of variables ($p=100$) compared to the sample size ($n=100$). There is one canonical variate pair and the canonical vectors are sparse. Only Sparse CCA and Robust Sparse CCA can be computed. 
The number of simulations for each setting is $M=500$.

\begin{table*}
\begin{center}
\caption{Simulation settings. In all settings ${ \bf \Sigma_{xx} }=\bf  {I}_p$ and ${ \bf \Sigma_{yy} }=\bf  {I}_q$. } \label{Settings}
\begin{tabular}{l|cccccccccc}
\hline
&&&&&&&&&&\\
Simulation setting & $n$ &&&$p$ &&& $q$  &&& $ { \bf \Sigma_{xy} }$  \\ 
&&&&&&&&&&\\ \hline
&&&&&&&&&&\\
Sparse Low-dimensional & 100 &&& 6 &&& 4 &&& $\begin{bmatrix}  0.9 & { \bf 0}_{ 1 \times 3} \\ { \bf 0}_{ 5 \times 1} & { \bf 0}_{ 5 \times 3}  \end{bmatrix}$ \\ 
&&&&&&&&&&\\
&&&&&&&&&&\\
Non-Sparse Low-dimensional & 250 &&& 12 &&& 8 &&&  $\begin{bmatrix}  0.2 & { \bf 0.1}_{ 1 \times 7}  \\ { \bf 0.1}_{ 11 \times 1} & { \bf 0.1}_{ 11 \times 7}  \\ \end{bmatrix}$ \\ 
 &&&&&&&&&&\\ 
 &&&&&&&&&&\\
Sparse High-dimensional & 100 &&& 100 &&& 4 &&&  $\begin{bmatrix}  { \bf 0.45}_{2 \times 2} & { \bf 0}_{2 \times 2}  \\ { \bf 0}_{98 \times 2} & { \bf 0}_{98 \times 98}   \end{bmatrix}$ \\ 
&&&&&&&&&&\\
\hline
\end{tabular}
\end{center}
\end{table*}

\bigskip

For each setting, the following sampling distributions are considered \citep{Branco05}
\begin{enumerate}
\item[(a)] \textit{No contamination.} We generate data matrices $\bf X$ and $\bf Y$ according to multivariate normal distributions $N_{p+q}({\bf 0}, {\bf \Sigma})$, with covariance matrices described in Table \ref{Settings}.
\item[(b)] \textit{Symmetric contamination.} 90\% of the data are generated from $N_{p+q}({\bf 0}, {\bf \Sigma})$, and 10\% of the data are generated from $N_{p+q}({\bf 0}, 9{\bf \Sigma})$.
\item[(c)] \textit{Asymmetric contamination.} 90\% of the data are generated from $N_{p+q}({\bf 0}, {\bf \Sigma})$, and 10\% of the observations equal the point $\text{tr}({\bf \Sigma}){\bf 1}^T$, where $\text{tr}({\bf \Sigma})$ is the trace of ${\bf \Sigma}$.
\end{enumerate}

\bigskip
\noindent
{\bf 4.1 Performance measures.} The estimators are evaluated on their estimation accuracy. We compute for each simulation run $m$, with $m=1,\ldots,M=500$, the angle $\theta^m({\bf \hat{A}}^m,{\bf A})$ between the subspace spanned by the estimated canonical vectors (contained in the columns of ${\bf \hat{A}}^m$) and the subspace spanned by the true canonical vectors (contained in the columns of $\bf A$).\footnote{The (minimum) angle $\theta(\widehat{\bf{A}},\bf{A})$ is computed as follows.
Compute the QR-decompositions $\widehat{\bf{A}}=\bf{Q}_{\widehat{\bf{A}}}\bf{R}_{\widehat{\bf{A}}}$ and $\bf{A}=Q_{\bf{A}}R_{\bf{A}}$. Next, compute the singular value decomposition of $Q_{\widehat{\bf{A}}}^{T}Q_{\bf{A}}=\bf{U}\bf{C}\bf{V}^{T}$. The matrix $\bf{C}$ is diagonal with elements $c_1,\ldots,c_r$.  The minimum angle is given by $\theta(\widehat{\bf{A}},\bf{A})=$$
\text{acos}(c_1)$.} Analogously for the matrix $\bf B$. The median angles are given by 
\begin{equation}
\theta_{\text{Med}}({\bf \hat{A}},{\bf A}) = \underset{1 \leq m \leq M}{\operatorname{med}}\theta^m({\bf \hat{A}}^m,{\bf A}) \text{\ \ \ and \ \ \ \ }
\theta_{\text{Med}}({\bf \hat{B}},{\bf B}) =\underset{1 \leq m \leq M}{\operatorname{med}}\theta^m({\bf \hat{B}}^m,{\bf B}). \nonumber
\end{equation}

For evaluating sparsity, we use the true positive rate and the true negative rate (e.g. \citealp{Rothman10})
\begin{gather}
\text{TPR}({\bf \hat{A}}^{m},{\bf A}) = \frac{ \# \{ (i,j):{\bf {\widehat{A_{ij}}}}^{m} \neq 0 \text{\ \ and \ }  {\bf {A_{ij}}} \neq 0   \}} { \# \{ (i,j):  {\bf {A_{ij}}} \neq 0         \}} \nonumber \\
\text{TNR}({\bf \hat{A}}^{m},{\bf A}) = \frac{ \# \{ (i,j):{\bf {\widehat{A_{ij}}}}^{m} = 0 \text{\ \ and \ }  {\bf {A_{ij}}} = 0   \}} { \# \{ (i,j):  {\bf {A_{ij}}} = 0         \}}. \nonumber 
\label{sparsityperformance} 
\end{gather}
Analogously for the matrix $\bf B$. A true positive is a coefficient that is non-zero in the true model, and is estimated as non-zero. A true negative is a coefficient that is zero in the true model, and is estimated as zero. Both the true positive rate and the true negative rate should be as high as possible for a sparse estimator.   

\bigskip
\noindent
{\bf 4.2 Results for the Sparse Low-dimensional design.} 
Simulation results for the estimator $\bf \hat{A}$ are in Table \ref{Rank1}. The results for the estimator $\bf \hat{B}$ are similar and are, therefore, omitted. In the scenario without contamination, the results are according to the expectations. 
The sparse estimators Sparse CCA and Robust Sparse CCA achieve a much better median estimation accuracy than the non-sparse estimators CCA and Robust CCA. As expected, a sparse method results in increased estimation accuracy when the true canonical vectors have a sparse structure.  
Looking at sparsity recognition performance,  Sparse CCA and Robust Sparse CCA perform equally good in retrieving the sparsity in the data generating process.

\begin{table}
\begin{center}
\caption{Results for the Sparse Low-dimensional design. Median of the angles between the space spanned by the true and estimated canonical vectors; median true positive rate and true negative rate are reported for each method. Averages of the angles are between parentheses.} \label{Rank1}
\footnotesize
\begin{tabular}{lccccccccccc} \hline
Method & \multicolumn{3}{c}{No contamination} && \multicolumn{3}{c}{Symmetric contamination} && \multicolumn{3}{c}{Asymmetric contamination} \\ 
& $\theta_{\text{Med}}({\bf \hat{A}},{\bf A})$ & TPR & TNR && $\theta_{\text{Med}}({\bf \hat{A}},{\bf A})$ & TPR & TNR && $\theta_{\text{Med}}({\bf \hat{A}},{\bf A})$ & TPR & TNR  \\ \hline
CCA 				& $\underset{(0.08)}{0.08}$ &      &       && $\underset{(0.18)}{0.16}$ &      &      &&  $\underset{(0.24)}{0.24}$ &      & \\
Robust CCA 			& $\underset{(0.11)}{0.10}$ &      &       && $\underset{(0.14)}{0.14}$ &      &      &&  $\underset{(0.14)}{0.14}$ &      & \\
Sparse CCA 			& $\underset{(0.02)}{0.00}$ & $\underset{(0.98)}{1.00}$  & $\underset{(0.97)}{1.00}$  && $\underset{(0.15)}{0.15}$ & $\underset{(1.00)}{1.00}$  & $\underset{(0.19)}{0.20}$ &&  $\underset{(0.19)}{0.19}$ & $\underset{(1.00)}{1.00}$  & $\underset{(0.10)}{0.01}$\\
Robust Sparse CCA 	& $\underset{(0.03)}{0.00}$ & $\underset{(0.99)}{1.00}$  & $\underset{(0.89)}{1.00}$   && $\underset{(0.03)}{0.01}$ & $\underset{(0.99)}{1.00}$  & $\underset{(0.84)}{0.80}$  &&  $\underset{(0.05)}{0.00}$ & $\underset{(0.98)}{1.00}$ & $\underset{(0.88)}{1.00}$ \\ \hline
\end{tabular}
\end{center}
\end{table}

In the contaminated simulation settings, we see that 
the robust estimators Robust Sparse CCA and Robust CCA maintain their accuracy under contamination. 
Robust Sparse CCA performs best. Robust Sparse CCA clearly outperforms Robust CCA: for instance, we have a median estimation accuracy of 0.01 (0.00) against 0.14 (0.14) for symmetric (asymmetric) contamination, see Table \ref{Rank1}.
The non-robust estimators CCA and Sparse CCA are clearly influenced by the introduction of outliers, as reflected by the much higher values of the median angles $\theta_{\text{Med}}({\bf \hat{A}},{\bf A})$. Sparse CCA now performs even worse than  Robust CCA. 
The non-robust estimators are most influenced in the more severe asymmetric contamination setting. 
Looking at sparsity recognition performance, Robust Sparse CCA method is hardly affected by the introduction of outliers. For the Sparse CCA method, especially the true negative rates is affected.

\bigskip

\noindent
{\bf 4.3 Results for the Non-sparse Low-dimensional design.} 
Summarizing results are in Table \ref{NonSparse}. Note that the true positive rate and true negative rate are omitted since the true canonical vectors are non-sparse. 
In the situation without contamination, 
all methods show similar performance. The price the sparse methods pay is a marginally decreased estimation accuracy, as measured by the median and average angle.
In the contaminated settings, the robust methods 
Robust Sparse CCA and Robust CCA perform best and show similar performance. 

\begin{table}
\begin{center}
\caption{Results for the Non-sparse Low-dimensional design. Median of the angles between the space spanned by the true and estimated canonical vectors are reported for each method. Averages of the angles are between parentheses.} \label{NonSparse}
\footnotesize
\begin{tabular}{lcccccccc} \hline
Method & \multicolumn{2}{c}{No contamination} && \multicolumn{2}{c}{Symmetric contamination} && \multicolumn{2}{c}{Asymmetric contamination} \\ 
 & \multicolumn{2}{c}{$\theta_{\text{Med}}({\bf \hat{A}},{\bf A})$} && \multicolumn{2}{c}{$\theta_{\text{Med}}({\bf \hat{A}},{\bf A})$} && \multicolumn{2}{c}{$\theta_{\text{Med}}({\bf \hat{A}},{\bf A})$} \\ \hline
CCA 				& \multicolumn{2}{c}{$\underset{(0.03)}{0.03}$ } && \multicolumn{2}{c}{$\underset{(0.18)}{0.09}$} && \multicolumn{2}{c}{$\underset{(0.33)}{0.10}$} \\

Robust CCA 				& \multicolumn{2}{c}{$\underset{(0.04)}{0.04}$ } && \multicolumn{2}{c}{$\underset{(0.04)}{0.04}$} && \multicolumn{2}{c}{$\underset{(0.05)}{0.04}$} \\

Sparse CCA 				& \multicolumn{2}{c}{$\underset{(0.03)}{0.03}$ } && \multicolumn{2}{c}{$\underset{(0.21)}{0.11}$} && \multicolumn{2}{c}{$\underset{(0.08)}{0.09}$} \\

Robust Sparse CCA 				& \multicolumn{2}{c}{$\underset{(0.04)}{0.04}$ } && \multicolumn{2}{c}{$\underset{(0.04)}{0.04}$} && \multicolumn{2}{c}{$\underset{(0.05)}{0.05}$} \\
 \hline
\end{tabular}
\end{center}
\end{table}

\bigskip

\noindent
{\bf 4.4 Results for Sparse High-dimensional design.} Summarizing results are in Table \ref{Highdim}.  In this high-dimensional setting, only Sparse CCA and Robust Sparse CCA can be performed. In the scenario without contamination, Sparse CCA performs best. Sparse CCA is, however, closely followed by Robust Sparse CCA both in terms of median estimation accuracy and sparsity recognition performance. 
When adding contamination, the performance of Sparse CCA gets distorted. The median estimation accuracy of Robust Sparse CCA is much better than the median estimation accuracy of Sparse CCA, especially in the more severe asymmetric contamination setting.

\begin{table}
\begin{center}
\caption{Results for Sparse High-dimensional design. Median of the angles  between the space spanned by the true and estimated canonical vectors; median true positive rate and true negative rate are reported for each method. Averages of the angles are between parentheses.} \label{Highdim}
\footnotesize
\begin{tabular}{lccccccccccc} \hline
Method & \multicolumn{3}{c}{No contamination} && \multicolumn{3}{c}{Symmetric contamination} && \multicolumn{3}{c}{Asymmetric contamination} \\ 
& $\theta_{\text{Med}}({\bf \hat{A}},{\bf A})$ & TPR & TNR && $\theta_{\text{Med}}({\bf \hat{A}},{\bf A})$ & TPR & TNR && $\theta_{\text{Med}}({\bf \hat{A}},{\bf A})$ & TPR & TNR  \\ \hline
Sparse CCA 			& $\underset{(0.13)}{0.06}$ & $\underset{(0.96)}{1.00}$ & $\underset{(0.97)}{0.98}$  && $\underset{(0.37)}{0.19}$ & $\underset{(0.92)}{1.00}$ & $\underset{(0.86)}{0.88}$ &&  $\underset{(0.34)}{0.33}$ & $\underset{(1.00)}{1.00}$ & $\underset{(0.98)}{1.00}$\\
Robust Sparse CCA 	& $\underset{(0.35)}{0.08}$ & $\underset{(0.83)}{1.00}$& $\underset{(0.98)}{0.99}$  && $\underset{(0.36)}{0.11}$ & $\underset{(0.84)}{1.00}$ & $\underset{(0.98)}{0.98}$ &&  $\underset{(0.33)}{0.08}$ & $\underset{(0.82)}{1.00}$ & $\underset{(0.99)}{1.00}$ \\ \hline
\end{tabular}
\end{center}
\end{table}
  
\bigskip

\noindent
Similar conclusions can be made when looking at the average angle as a measure of estimation accuracy (reported in Tables \ref{Rank1} to \ref{Highdim} between parentheses). In sum, Robust Sparse CCA shows the best overall performance in this simulation study. Robust Sparse CCA combines both robustness and sparsity. Therefore, it performs best in sparse settings where contamination is present. In sparse non-contaminated settings, Robust Sparse CCA shows competitive performance to its main competitor Sparse CCA. In contaminated non-sparse settings, Robust Sparse CCA shows competitive performance to its main competitor Robust CCA.

\section{Applications}
\noindent
{\bf 5.1 Evaporation data set.} As a first example, we use a data set from \cite{Freund79}. Ten variables (maximum, minimum and average soil temperature; maximum, minimum and average air temperature; maximum, minimum and average daily relative humidity; and total wind) have been measured on 46 consecutive days from June 6 until July 21. The aim is to find and quantify the relations between the soil temperature variables and the remaining variables.

\begin{figure}
\begin{center}
\includegraphics[width=10cm]{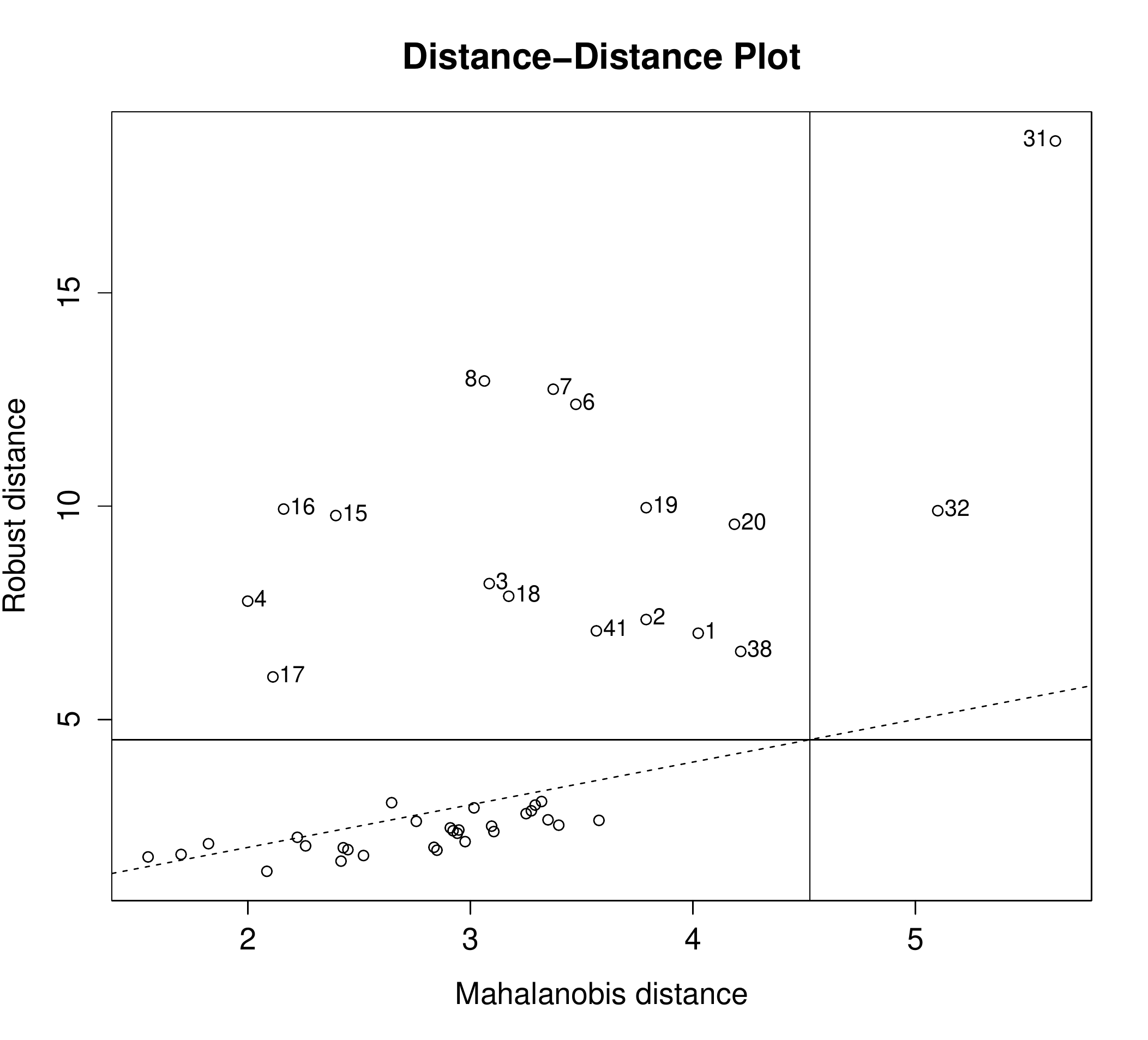} 
\caption{Distance-Distance Plot for the Evaporation data} \label{DDplot_Evaporation}
\end{center}
\end{figure}

As a first inspection of the data, we use the Distance-Distance plot \citep{Rousseeuw90} in Figure \ref{DDplot_Evaporation}. The Distance-distance plot displays the robust distances versus the Mahalanobis distances. The vertical and horizontal lines are drawn at values equal to the square root of the 97.5\% quantile of a chi-squared distribution with 10 degrees of freedom. Points beyond those lines would be considered as outliers. The Distance-Distance plot reveals some outliers: objects 31 and 32, for example, are extreme outliers. This suggests the need for a robust CCA method.

We use the maximum eigenvalue ration criterion (e.g. \citealp{An13}), as discussed in Section 3, to decide on the number of canonical variate pairs to extract. For all CCA methods two canonical variate pairs are extracted. To compare the performance of the CCA approaches, we perform a leave-one-out cross-validation exercise and compute the cross-validation score
\begin{equation}
CV = \frac{1}{h} \sum_{i=1}^{h} || {\bf \widehat{A}}^{T}_{-i}{\bf x}_i  - {\bf \widehat{B}}^{T}_{-i}{\bf y}_i ||^2, \nonumber
\end{equation}
where ${\bf \widehat{A}}^{T}_{-i}$ and ${\bf \widehat{B}}^{T}_{-i}$  contain the estimated canonical vectors when the $i^{th}$ observation is left out of the estimation sample and $h=\lfloor n\alpha\rfloor$, with $\alpha=1$ (0\% Trimming) or $\alpha=0.9$ (10\% Trimming). We use trimming to eliminate the effect of outliers in the cross-validation score. The method that achieves the lowest cross-validation score has the best out-of-sample performance.  Table \ref{CVscore_Evaporation} reports the cross-validation scores for all methods.  Robust Sparse CCA achieves the best cross-validation score. 

\linespread{1.4}
\begin{table}
\caption{Evaporation data set: Cross-validation score for each method.} \label{CVscore_Evaporation}
\begin{center}
\begin{tabular}{lccccc}
  \hline
Method              &&& CV-score &&  CV-score\\ 
             &&& 0\% Trimming &&  10\% Trimming\\ \hline
CCA 	            &&& 1.48 && 0.98 \\
Robust CCA 			&&& 1.26 && 0.73 \\
Sparse CCA 			&&& 1.63 &&  0.68 \\
Robust Sparse CCA 	&&& 0.90 && 0.66 \\ \hline
\end{tabular}
\end{center}
\end{table}
\linespread{1.6}

\linespread{1.4}
\begin{table}
\caption{Evaporation data set: Estimated canonical vectors using Robust CCA and Robust Sparse CCA.} \label{CanVectors}
\small
\begin{center}
\begin{tabular}{llcccccccccc}
  \hline
 && \multicolumn{4}{r}{Robust CCA} &&&& \multicolumn{3}{c}{Robust Sparse CCA } \\ 
\multicolumn{4}{l}{Variables $\backslash$ Canonical Vectors}  & 1 & 2 &&&&& 1 & 2 \\\hline
\multicolumn{2}{l}{\textit{First data set}} &&&&&&&&&& \\
 \multicolumn{3}{l}{MAXST: Max. daily soil temperature} &&  -0.21 & -0.73 &&&&& 0 & 0.76 \\ 
 \multicolumn{3}{l}{MINST: Min. daily soil temperature}&&   -0.02 & 0.67 &&&&& 0 & -0.63 \\ 
 \multicolumn{3}{l}{AVST: Avg. daily soil temperature}&&  0.98 & 0.13 &&&&& 1 & -0.15 \\ 
 &&&&&&&&&&& \\
\multicolumn{2}{l}{\textit{Second data set}} &&&&&&&&&& \\
 \multicolumn{3}{l}{MAXAT: Max. daily air temperature} &&    0.61 & 0.01  &&&&& 0.94 & 0\\ 
 \multicolumn{3}{l}{MINAT: Min. daily air temperature}  &&   0.53 & 0.75 &&&&&  0.15 & -0.84 \\ 
 \multicolumn{3}{l}{AVAT: Avg. daily air temperature}  && 0.08 & -0.05 &&&&&  0.17 & 0 \\ 
 \multicolumn{3}{l}{MAXH: Max. daily relative humidity}  &&  -0.11 & 0.05  &&&&& 0 & 0 \\ 
 \multicolumn{3}{l}{MINH: Min. daily relative humidity}  &&  0.16 & 0.54 &&&&& 0 & -0.53 \\ 
 \multicolumn{3}{l}{AVH: Avg. daily relative humidity}  && -0.43 & 0.33 &&&&& -0.24 & 0 \\ 
 \multicolumn{3}{l}{WIND: Total wind, measured in miles per day}  &&  -0.35 & 0  &&&&& 0 & 0 \\ \hline 
\textit{Canonical correlations}  &&&& 0.89 & 0.59 &&&&& 0.87 & 0.48   \\ \hline 
\end{tabular}
\end{center}
\end{table}
\linespread{1.6}

Table \ref{CanVectors} shows the estimated canonical vectors for the Robust CCA and Robust Sparse CCA method. 
By adding the penatly term, the number of non-zero coefficients is reduced from 20 to 10, improving interpretability of the canonical vectors. The price to pay for the sparseness is a slight decrease in the estimated canonical correlations (computed using the bivariate MCD estimator, see Section 3): they drop from 0.89 to 0.87 for the first one, and from 0.59 to 0.48 for the second canonical correlation. 
We find this decrease acceptable, given the gained sparsity in the canonical vectors. The sparse structure of the canonical vectors facilitates interpretation. The first canonical variate in the soil temperature data set, for instance, is uniquely determined by the variable AVST. 
\bigskip

\noindent
{\bf 5.2 Nutrimouse data set.} As a second example, we use the nutrimouse data set  (publicly available in the R package \verb+CCA;+ \citealp{Gonzalez08}). Two sets of variables, i.e. gene expressions and fatty acids, are available for 40 mice. The first set contains expressions of 120 genes measured in liver cells. The second set of variables contains concentrations of 21 hepatic fatty acids (FA). Furthermore, the mice are classified according to two factors: genotypes (wild-type and PPAR$\alpha$ deficient mice) and diets (reference diet ``REF", hydrogenated coconut oil diet ``COC", sunflower oil diet ``SUN", linseed oil diet ``LIN" and fish oil diet ``FISH"). More details on how the data were obtained can be found in \cite{Martin07}. The aim is to identify a small set of genes which are correlated with the fatty acids.

In this  data set, the number of experimental units (i.e. data on 40 mice) is smaller than the number of variables (i.e. 120 genes). Therefore, standard CCA nor robust CCA can be performed. \cite{Gonzalez08} use the Canonical Ridge. The Canonical Ridge performs regularization, but the canonical vectors produced by the Canonical Ridge are not sparse. Robust Sparse CCA  and Sparse CCA, can also be applied in this high-dimensional setting and produce -  more interpretable - sparse canonical vectors. 

\linespread{1.4}
\begin{table}
\caption{Nutrimouse data set: Cross-validation score for each method.} \label{CVscore_Nutrimouse}
\begin{center}
\begin{tabular}{lccccc}
  \hline
Method              &&& CV-score && CV-score\\ 
             		&&&  0\% Trimming && 10\% Trimming \\ \hline
Sparse CCA 			&&& 190.89  &&  114.42\\
Robust Sparse CCA 	&&& 78.57   && 37.10 \\ \hline
\end{tabular}
\end{center}
\end{table}
\linespread{1.6}

We compare the performance of the Robust Sparse CCA method to the Sparse CCA method. For both methods, two canonical variate pairs are extracted using the maximum eigenvalue ration criterion from Section 3. We perform the same leave-one-out cross-validation exercise as described in Section 5.1. Results are reported in Table \ref{CVscore_Nutrimouse}. The Robust Sparse CCA method outperforms the Sparse CCA method. The cross-validation scores are reduced by about 60\% when using the robust method.

\begin{figure}
\begin{center}
\includegraphics[width=14cm]{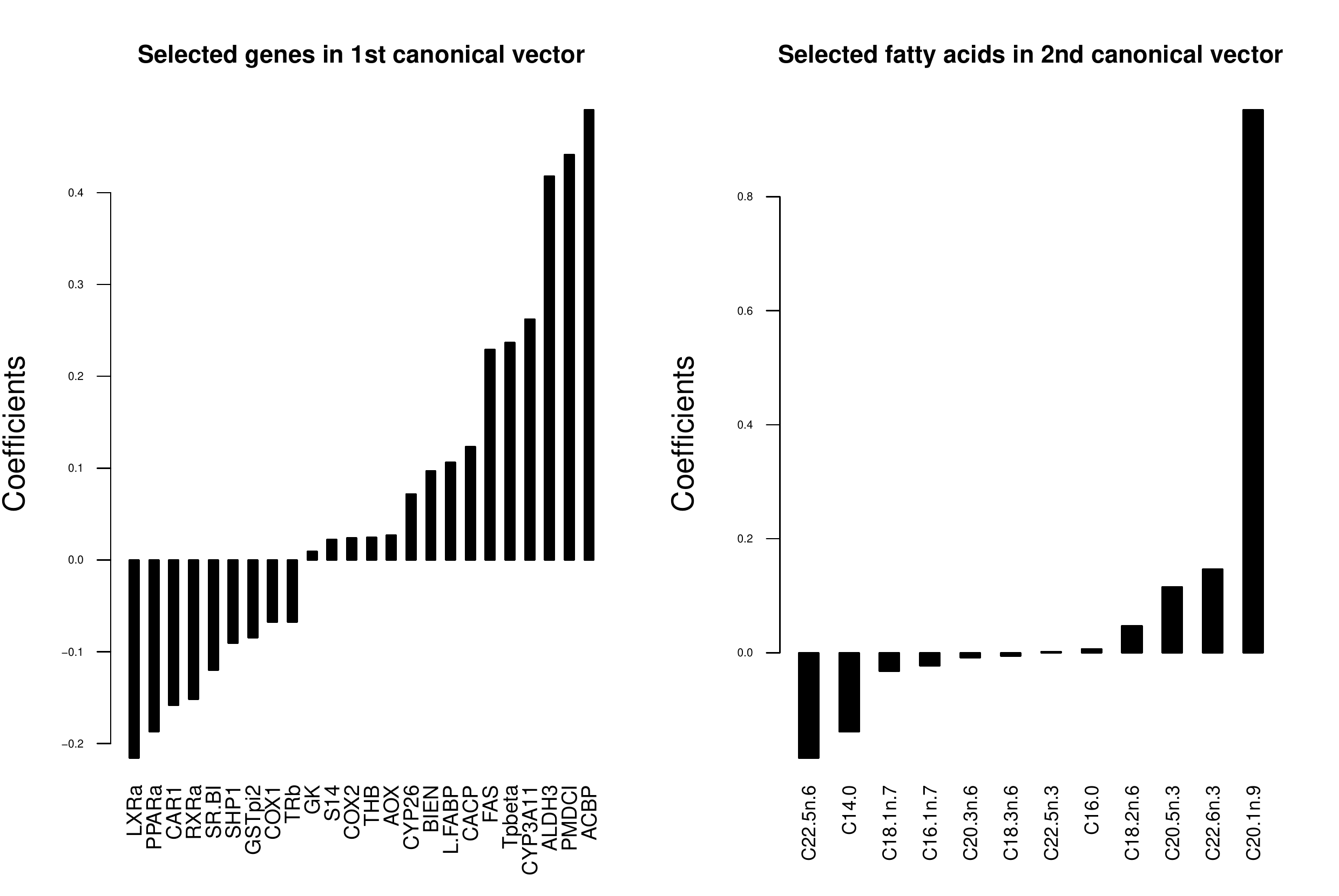} 
\caption{\small Coefficients of selected genes (left) and coefficients of selected fatty acids (right). } \label{Nutrimouse}
\end{center}
\end{figure} 


Next, we discuss the estimated canonical vectors obtained using the Robust Sparse CCA method. The most informative biological conclusions can be drawn from (i) the selected genes in the first canonical vector and (ii) the selected fatty acids in the second canonical vector (for further motivation see, \citealp{Martin07,Gonzalez08}). Therefore, we focus on these results. First, the left panel of Figure \ref{Nutrimouse} displays the coefficients of the selected genes, i.e. those genes with non-zero estimated coefficients, in the first canonical vector. 24 out of 120 variables are selected. The solution  is very sparse, facilitating interpretation. Two important genes are $\verb|Cyp3a11|$  and $\verb|CAR1|$.  \cite{Martin07} find a consistent reduction of $\verb|Cyp3a11|$ in PPAR$\alpha$ livers on the one hand, and an overexpression of  $\verb|CAR1|$ on the other hand. Both genes are selected and have among the highest (absolute) coefficients. Second, we focus on the fatty acids and discuss the right panel of Figure \ref{Nutrimouse}. 
The right panel displays the coefficients of the selected fatty acids in the second canonical vector. 12 out of 21 fatty acid variables are selected. The 
fatty acids are related to the effect of certain diets used in this experiment. The mice are classified according to a specific diet. Five diets are used, which contain specific concentration levels of the fatty acids $\verb|C22:6n-3|$, $\verb|C22:5n-3|$, $\verb|C22:5n-6|$, $\verb|C22:4n-3|$ and $\verb|C20:5n-3|$ respectively \citep{Martin07}.  Looking at the selected fatty acids in the second canonical vector, we see that four out of these five variables  are selected and have among the highest (absolute) coefficients.

In sum, the strong genotype effect observed through the first canonical variate and the strong diet effect observed through the second canonical variate is in accordance with conclusions drawn in \cite{Martin07}.

\section{Conclusion}
Sparse Canonical Correlation Analysis delivers interpretable canonical vectors, with some of its elements estimated as exactly zero. Robust Sparse CCA retains this advantage, while at the same ensuring that the estimation of the canonical vectors is not affected by outlying observations. 

The canonical vectors are given by the eigenvectors of two particular matrices, see for instance \citeauthor{Johnson98} (\citeyear{Johnson98}, Chapter 10). Typically, the canonical vectors are estimated by taking the sample versions of those covariance matrices and computing the corresponding eigenvectors. One could think of estimating those covariance matrices with an estimator that is robust and sparse at the same time, and then, to compute the eigenvectors. This approach, however, would results in canonical vectors being non-sparse. To circumvent this pitfall, we reformulate the CCA problem in a regression framework. We use the sparse LTS estimator of \cite{Alfons13} to obtain sparse canonical vectors that are resistant to outlying observations. 

A simulation study and two empirical examples show the advantages of  the Robust Sparse CCA method over its natural benchmarks. In contaminated settings, Robust Sparse CCA outperforms standard CCA and  Sparse CCA.  Robust Sparse CCA has two important advantages over Robust CCA. First, Robust Sparse CCA improves model interpretation since only a limited number of variables, those corresponding to the non-zero elements of the canonical vectors, enter the estimated canonical variates (cfr. evaporation application). Second, if the number of variables approaches the sample size, the estimation precision of Robust CCA suffers. If the number of variables exceeds the sample size, Robust CCA can not even be performed. Due to the regularization performed by Robust Sparse CCA, Robust Sparse CCA can still be computed (cfr. nutrimouse application).

Several questions are left for future research. One could think of a joint selection criterion for the number of canonical variate pairs and the sparsity parameter. This would, however, increase computation time substantially. To induce sparsity in the canonical vectors, we use a Lasso penalty. Other penalty functions such as the Adaptive Lasso \citep{Zou06} could be considered. The Adaptive Lasso is consistent for variable selection, whereas the Lasso is not. Furthermore, we use a regularized version of the LTS estimator. One could also use a regularized version of the S-estimator or the MM-estimator to increase efficiency. Up to our knowledge, however, the sparse LTS is the only robust sparse regression estimator for which efficient code \citep{robustHD} is available. 

\section*{Acknowledgments}
\noindent Financial support from the FWO (Research Foundation Flanders) is gratefully acknowledged (FWO, contract number 11N9913N).

\bibliographystyle{asa}
\bibliography{robustsparsecca_ref}

\end{document}